# The effect of pressure on hydrogen solubility in Zircaloy-4

H.E. Weekes[,2], D. Dye[1*], J.E. Proctor[3,4], D. Smith[3,4], C. Simionescu[3], T.J. Prior[6], and M.R. Wenman[1]

[1]Department of Materials, Royal School of Mines, Imperial College, London, SW7 2BP, UK
[2]National Nuclear Laboratory, Culham Science Centre, Building D5, Abingdon, Oxfordshire, OX14 3DB, UK
[3]School of Computing, Science & Engineering, University of Salford, Salford, M5 4WT, UK
[4]Department of Physics & Mathematics, University of Hull, Hull, HU6 7RX, UK
[6]Department of Chemistry, School of Mathematical and Physical Science, University of Hull, Hull, HU6 7RX, UK
* corresponding author: david.dye@imperial.ac.uk

**Abstract**

The effect of pressure on the room temperature solubility of hydrogen in Zircaloy-4 was examined using synchrotron X-ray diffraction on small ground flake samples in a diamond anvil cell at pressures up to 20.9 GPa. Different combinations of hydrogen level/state in the sample and of pressure transmitting medium were examined; in all three experiments, it could be concluded that pressure resulted in the dissolution of $\delta$ hydrides and that interstitial hydrogen seemingly retards the formation of $\omega$ Zr. A pressure of around 9 GPa was required to halve the hydride fraction. These results imply that the effect of pressure is thermodynamically analogous to that of increasing temperature, but that the effect is small. The results are consistent with the volume per Zr atom of the $\alpha$, $\delta$ and $\omega$ phases, with the bulk moduli of $\alpha$ and $\delta$, and with previous measurements of the hydrogen site molar volumes in the $\alpha$ and $\delta$ phases. The results are interpreted in terms of their implication for our understanding of the driving forces for hydride precipitation at crack tips, which are in a region of hydrostatic tensile stress on the order of 1.5 GPa.

1. Introduction

Zirconium alloys used in water cooled nuclear reactors are susceptible to hydrogen pick-up in service leading to embrittlement and in some cases a time dependent failure mechanism known as delayed hydride cracking (DHC). The mechanism of DHC requires a stress raiser, such as a notch, which then attracts hydrogen dissolved in the bulk zirconium metal, eventually resulting in the precipitation of a brittle hydride phase. The macro-hydride precipitate (made up of stacks of micro-hydrides) then grows until it reaches a critical size and fractures. At this point the new crack tip attracts more hydrogen and the process repeats itself, ultimately leading to failure of components. For decades the mechanisms for DHC have been debated from the original proposals of Dutton and Puls [1-2], who first suggested that tensile stress was a critical feature of DHC, by both reducing the chemical potential of hydrogen and through the reduction of the solubility limit of hydrogen in zirconium, through to more recent models of Kim *et al.* [3-5] and McRae *et al.* [6]. These recent models were labelled in the literature by McRae as the precipitation first model (PFM) and the diffusion first model (DFM), respectively, and much debate occurred [7-11]. The subject of this work is not the details of these mechanisms, but on one key difference between DFM and PFM models that allows them to be tested experimentally. The PFM model requires that tensile stress, created by the presence of a crack, reduces the solubility limit of hydrogen in $\alpha$-zirconium, by a substantial amount. This results in the precipitation of hydride as a first step of DHC. In contrast, the DFM model does not require any reduction in hydrogen solubility with tensile stress, as it is based on the diffusion of hydrogen to the crack tip, due to reduction of the chemical potential of hydrogen in the presence of a tensile stress field.

In between the development of the early and more recent mechanistic models of DHC, work was done to assess whether the solubility of hydrogen did indeed change significantly with applied tensile stress and, if so, by how much. Coleman's experiments on stressed pre-cracked Zr-2.5%Nb samples showed no detectable differences in the stressed and unstressed solubility limit but were limited in the testable range of stress by yielding of the material [12]. MacEwen *et al.* subsequently showed, through careful neutron measurements of the lattice parameter of $\alpha$-zirconium, that the difference in the molar volumes between hydrogen in the hydride phase and dissolved hydrogen in the zirconium matrix, was smaller than originally thought [13]. From this they estimated that the reduction in solubility due to a tensile stress of 500 MPa would be less than 3% at 570 K and hence within the scatter of experiments conducted. Since this experiment, whilst debate on DHC mechanisms continues, little has been done to verify the results of MacEwen *et al.* although recently, Blackmur *et al.* have also observed a similar dilation using synchrotron X-ray diffraction, suggested to be associated with hydrogen [14], and we and others have examined the effect of stress on hydride cycling [15,16] and hydride micromechanics [17].

Here, to avoid the limitations on the stress range obtainable in tension tests, encountered by previous workers, compressive strain was applied instead with the hypothesis that the effect on hydrogen solubility should be the same as in tension but simply in a reverse sense. This was achieved by pressurising samples under quasi-hydrostatic conditions inside a diamond anvil cell (DAC) allowing the examination of a much greater range of stress than achievable in tension tests from close to zero up to ~20 GPa. From this it is possible to study whether compressive strain causes hydrogen to go into solution in Zircaloy more readily, or precipitate out of solution. Investigation into the effect of interstitial hydrogen in Zircaloy on the stability of the α phase relative to the high pressure ω phase (hexagonal) was also carried out. Pure zirconium transforms from α phase to ω phase at ambient temperature at around 3.5 GPa pressure [18] although a range of transition pressures have been observed, sensitive to impurities [19].

In this study we measure, in situ using synchrotron X-ray diffraction, the effect of stress on the solubility of hydrogen in $\alpha$-zirconium by applying quasi-hydrostatic compressive stress to Zircaloy-4 samples with hydrogen dissolved in the α-Zr structure, as well as hydrogen that has caused the precipitation of the δ-hydride phase. We also examine whether hydrogen can be induced to enter Zircaloy by the application of compressive strain.

## 2. Experimental

Experiments under high hydrostatic pressure were conducted in custom-constructed donut-type (i.e short piston-cylinder) DACs equipped with rhenium and stainless steel gaskets, and diamonds with 450 μm or 250 μm culets at room temperature. Zircaloy-4 samples made from hot-rolled plate and sourced ultimately from ATI Wah Chang (see Table 1 for composition) had hydrogen added electrochemically. It should be noted that the measured Sn content lies slightly outside the specification limit of 1.7 wt%, but probably within the ICP-OES uncertainty for the measuring laboratory. The electrochemical hydrogen addition was done by welding electrodes of zirconium and platinum to the sample and a platinum wire mesh respectively and connecting them to a power supply. The zirconium sample acted as the cathode while the mesh was the anode. Samples were immersed in an electrolytic solution of diluted $H_2SO_4$ ($H_2SO_4:H_2O \rightarrow 2:100$), at a temperature of 65˚C (±5˚C) with an applied current density of 2 kA m$^{-2}$. Samples formed a (relatively) uniform surface hydride layer of approximately 60 μm thickness, equivalent (if dissolved into the entire sample at temperature) to a hydrogen concentration of circa 300 wppm; this is taken as the H content. The sample was then diffusion annealed at 300 ˚C in an encapsulated inert gas (Ar) filled quartz tube to prevent oxidation in order to diffuse hydrogen throughout the sample; hydrides then precipitated on cooling leaving the metal matrix saturated with hydrogen at the room temperature solubility limit of <1wppm [20]. Full details of the charging procedure are provided in Ref. [21].

Table 1. Chemical composition (measured, ICP-OES, Incotest Hereford) of the hot-rolled and recrystallized Zircaloy-4 plate utilised.

| Element | Cr | Fe | N | Nb | Ni | O | Sn | Zr |
|---|---|---|---|---|---|---|---|---|
| Weight % | 0.12 | 0.22 | 0.003 | 0.01 | <0.01 | 0.13 | 1.78 | Bal. |

Both as-received and pre-hydrided samples were ground to fine powders with a diameter < 200 μm. Flakes from the powder were manually selected and placed in the sample chamber in the DAC gasket. A ruby microcrystal was also placed in the sample chamber to enable measurement of pressure using the established ruby fluorescence technique [22].

Samples were pressurized in both a hydrogen pressure transmitting medium (PTM) and an inert PTM (4:1 methanol-ethanol solution). Hydrogen was loaded by closing the DAC under 2 kbar of hydrogen pressure in a custom-designed gas loader apparatus [23,24], and 4:1 methanol-ethanol solution was loaded by placing a droplet of the solution over the sample chamber and closing the DAC before the solution was able to evaporate. The time between pressure changes and measurements were in excess of 10minutes, providing ample time for hydrogen diffusion at room temperature. Methanol-ethanol is a good choice of PTM as it has been shown to be inert at room temperature, similar to argon, and to achieve quasi-hydrostatic conditions through gradual solidification into a disordered structure rather than rapid crystallization [25]. It is worth noting that hydrogen is not a gas in these experiments but is expected to transform to a rigid liquid at around 0.2 GPa [26] and solidifies at 5.5 GPa [27] at 300 K.

The hydrogen compressibility in the pressure range of these experiments is greatly reduced compared to the gas phase [28,29]. However, its ability to diffuse into different materials is not. If anything, hydrogen becomes even more reactive and diffusive under high hydrostatic pressure in the solid state [30-33].

X-ray diffraction (XRD) patterns were collected at the Diamond Light Source using 30 keV monochromatic X-rays and a MAR345 image plate area detector. The FIT2D software package was used to calibrate diffraction parameters and integrate the diffraction rings to 1-dimensional intensity-2θ plots [34]. The peak positions in the 1-dimensional plots were determined using multi-peak fitting in IgorPro with the pseudo-Voigt function, a convolution of the Gaussian and Lorentzian functions. The (0002), $\{10\bar{1}1\}$ and $\{11\bar{2}0\}$ diffraction peaks from the $\alpha_{Zr}$ phase were fitted individually using a single-peak function whilst the $\{10\bar{1}0\}$ peak from the $\alpha_{Zr}$ phase was fitted using a multi-peak fit due to its proximity to the $\{111\}$ peak from the δ-hydride phase.

The $a_{Zr}$ lattice constant was calculated from the fitted $d$-spacings for the $\{10\bar{1}0\}$ and $\{11\bar{2}0\}$ reflections and the average taken of these values. The $c_{Zr}$ lattice constant was then calculated from the fitted $d$-spacings for the (0002) and $\{10\bar{1}1\}$ reflections and the average taken of these values. The unit cell volume was then calculated from these lattice parameters.

We conducted our own measurement of the unit cell volume and lattice parameters of as-received Zircaloy-4 at ambient conditions and obtained values closely matching those available in the literature [35]. The unit cell volume and lattice parameters are shown in Table 2. The lattice constant $a_{ZrH}$ and unit cell volume for the hydride phase were hence calculated from the fitted $d$-spacing for the $\{111\}$ peak, the only peak observed from the $\delta_{ZrH}$ phase. Other studies (Ref. [21] and citations therein) have shown this to be by far the most intense peak from the δ phase, as expected.

Table 2. Comparison between the α-Zr lattice parameters measured at ambient conditions in the present work and those obtained from Ref [31].

| Source/lattice parameter | a (Å) | c (Å) | c/a |
|---|---|---|---|
| This work | 3.2310 | 5.1515 | 1.5944 |
| Steuwer et al. [31] | 3.2276 | 5.1516 | 1.5961 |

## 3. Results

Three high pressure experiments were conducted, at room temperature, the results of which will be described in turn.

### 3.1 Compression in the absence of hydrogen

An inert PTM, 4:1 methanol-ethanol solution was used to compress a sample of Zircaloy-4 that had been electrochemically charged. In this case, the hydrogen was found to be all in solution in the *hcp* α-Zr matrix host causing a modest lattice expansion, i.e. the δ-ZrH phase was not observed, suggesting that the flakes selected contained no hydrides. This sample would therefore only be expected to contain the room temperature solubility limit of hydrogen (<1wppm [20]), for contrast with the hydride-containing samples examined subsequently. In this experiment precipitation of the δ hydride phase was not observed at any point, however a partial phase transformation from the α-Zr phase to the ω-Zr phase was observed (commencing at 19 GPa, just below the highest pressure reached of 20.9 GPa). Selected diffraction patterns are shown in Figure 1. The reduction in intensity of the $(0002)_\alpha$ peak and increase in the $\{11\bar{2}0\}_\omega$ is consistent with the Silcock orientation relationship $(0002)_\alpha \parallel \{11\bar{2}0\}_\omega$ [36].

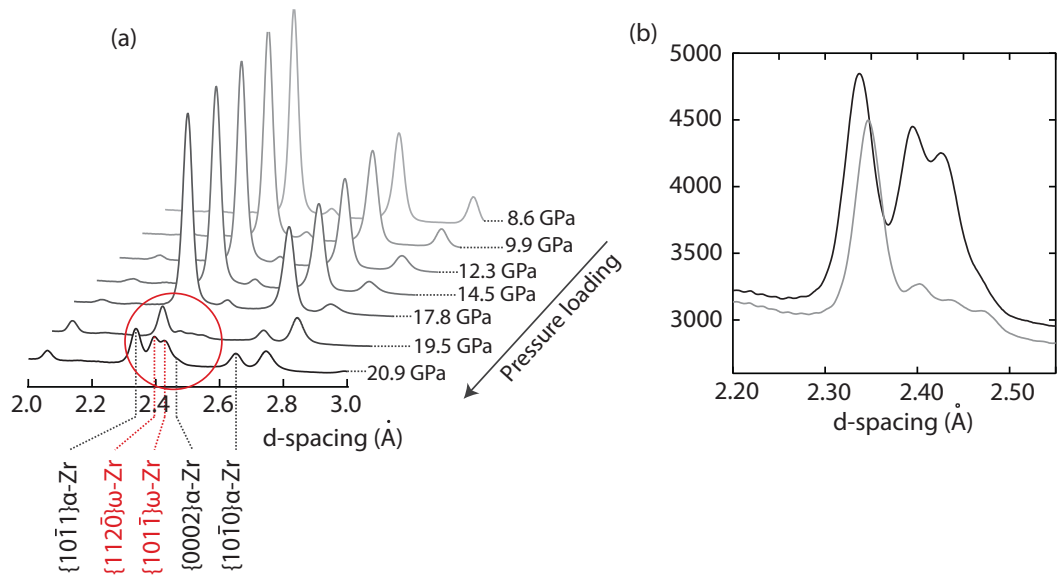

Figure 1. (colour online). (a) Evolution in the diffraction peaks from the α and ω phases of Zircaloy with compression. y-axis shows intensity (au). (b) shows a magnified view of the rapid appearance of the ω peaks around 2.4 Å from 19.5 to 20.9 GPa.

### 3.2 Compression of as-received Zircaloy-4 to 13.7 GPa pressure in hydrogen PTM

Upon closure of the DAC in the hydrogen atmosphere, the pressure in an as-received, unhydrided sample was increased to 6.6 GPa (the hydrogen is now solid [27]) and the first X-ray diffraction pattern was taken. At this pressure the expected diffraction peaks from the $\alpha_{Zr}$ phase were present. An additional peak could be observed, overlapping with the $\{10\bar{1}0\}_{\alpha\text{-Zr}}$ peak (Figure 2), which we attribute to the $\{111\}_{\delta\text{-ZrH}}$ peak of the hydride phase. Therefore, initial loading to 6.6 GPa caused enough hydrogen to adsorb and penetrate into the zirconium matrix to exceed the solubility limit in the outer layers of the sample and cause the precipitation of the δ-hydride. The δ phase peak only appeared as a broad shoulder, consistent with size and strain broadening of the hydride phase, but its appearance is consistent with that observed in Ref. [35].

Figure 2 shows the fitting of the relevant peaks at a range of pressures, and the intensity of the $\{111\}_\delta$ peak relative to that of the $\{10\bar{1}0\}_{\alpha-Zr}$ peak (Figure 2d). Upon further compression to 13.7 GPa the intensity of the $\{111\}_\delta$ peak decreased significantly. Diffraction patterns were then collected upon decompression to 3.7 GPa and as the sample was decompressed the intensity of the $\{111\}_{\delta-ZrH}$ peak increased. Therefore, hydride dissolution appeared to occur on the application of pressure, and re-precipitation on unloading. If equilibrium is assumed, this implies that the hydrogen content of the sample increased during the loading-unloading cycle, due to the hydrogen medium used.

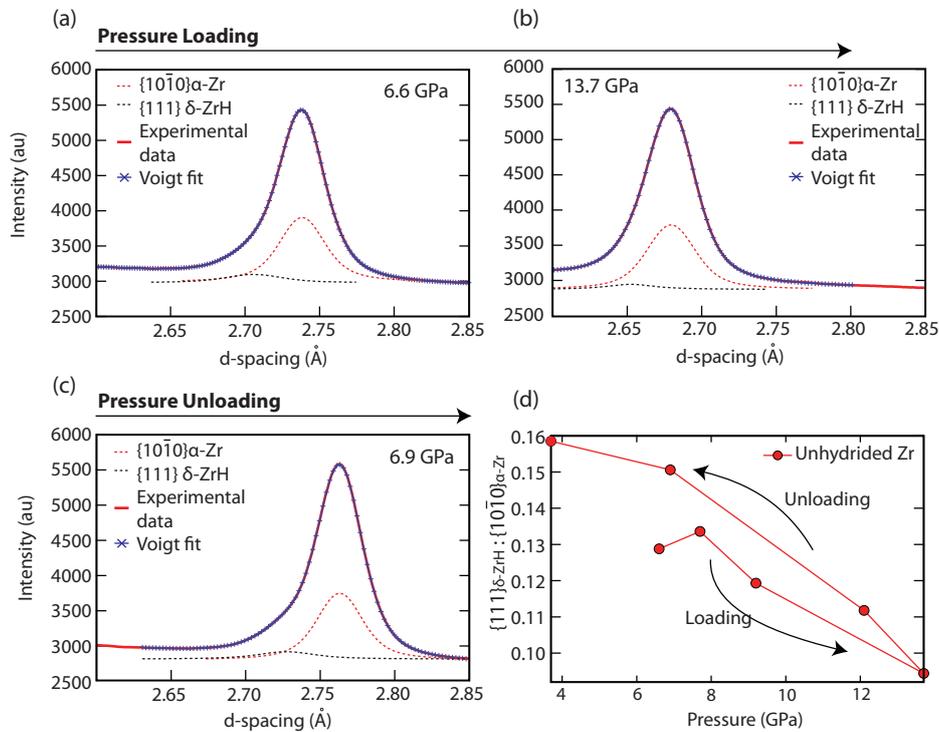

Figure 2 (colour online). (a-c) Variation in the δ-hydride $\{111\}$ peak intensity relative to the neighbouring $\{10\bar{1}0\}$ α-Zr peak on compression and decompression in an as-received Zircaloy-4 sample, compressed in hydrogen PTM. (d) Change in δ $\{111\}$ / α $\{10\bar{1}0\}$ peak intensity ratio upon compression and decompression.

## 3.3 Compression of electrochemically hydrided Zircaloy-4 with δ already present in hydrogen PTM

A sample of electrochemically hydrided Zircaloy-4, in which the δ-hydride phase was already present along with the α phase, was compressed to 20 GPa pressure in a hydrogen PTM. An overview of the diffraction spectra obtained is shown in Figure 3. In this experiment we also observed the appearance of diffraction peaks characteristic of the ω phase of Zircaloy, from 17 GPa onwards upon compression (not shown). These $ω_{Zr}$ reflections persisted during the unloading phase, at 12.7 through to 1.4 GPa, Figure 3.

The evolution of the peaks around 2.75 Å during the loading cycle is shown in Figures 4-5. At the initial, lowest pressure data point at 7.8 GPa, the $\{111\}_{\delta-ZrH}$ peak was observed as a low intensity broad shoulder to the left of the $\{10\bar{1}0\}_{\alpha-Zr}$ diffraction peak. Similar to the observations with the as-received Zircaloy, the $\{111\}_{\delta-ZrH}$ peak gradually decreased in intensity upon compression, becoming unobservable above ~14 GPa, Figure 4. The peak then re-appeared at a pressure of ~4.4 GPa upon unloading, increasing significantly in intensity upon further decompression, Figure 5. At the lowest load (1.4 GPa), the intensity of the $\{111\}_{\delta-ZrH}$ peak finally drops again.

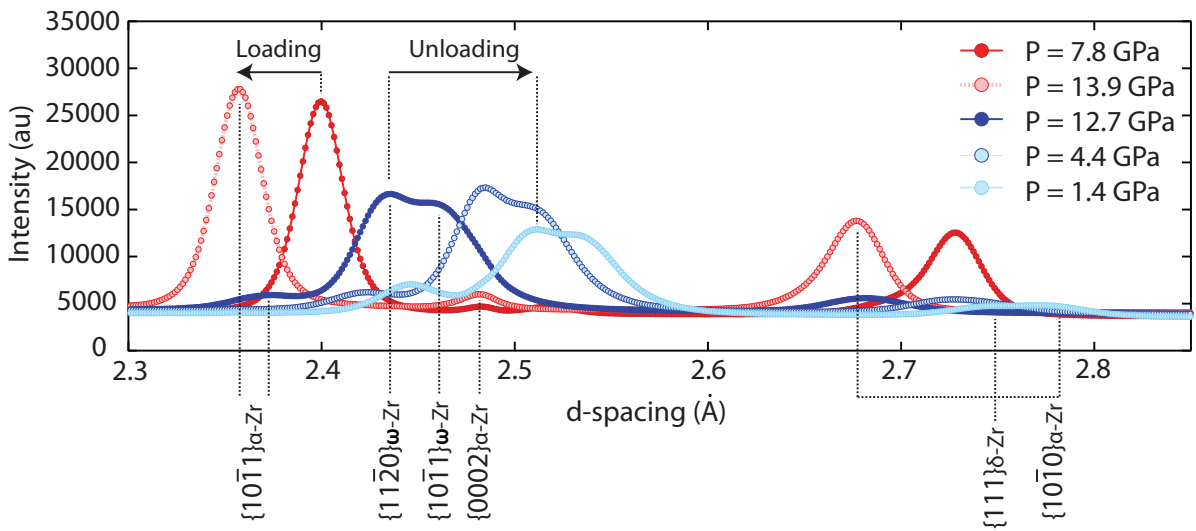

Figure 3 (colour online) (a) Evolution of the diffraction from a pre-hydrided Zircaloy-4 sample upon compression (red), and decompression (blue) in a hydrogen PTM. On loading, the $\{10\bar{1}1\}_{\alpha\text{-Zr}}$ peak moves to lower d-spacing. At peak pressure and on unloading, the $\{11\bar{2}0\}_{\omega\text{-Zr}}$ and $\{10\bar{1}1\}_{\omega\text{-Zr}}$ doublet appears. The $\{111\}_{\delta\text{-ZrH}}$ peak can also be observed.

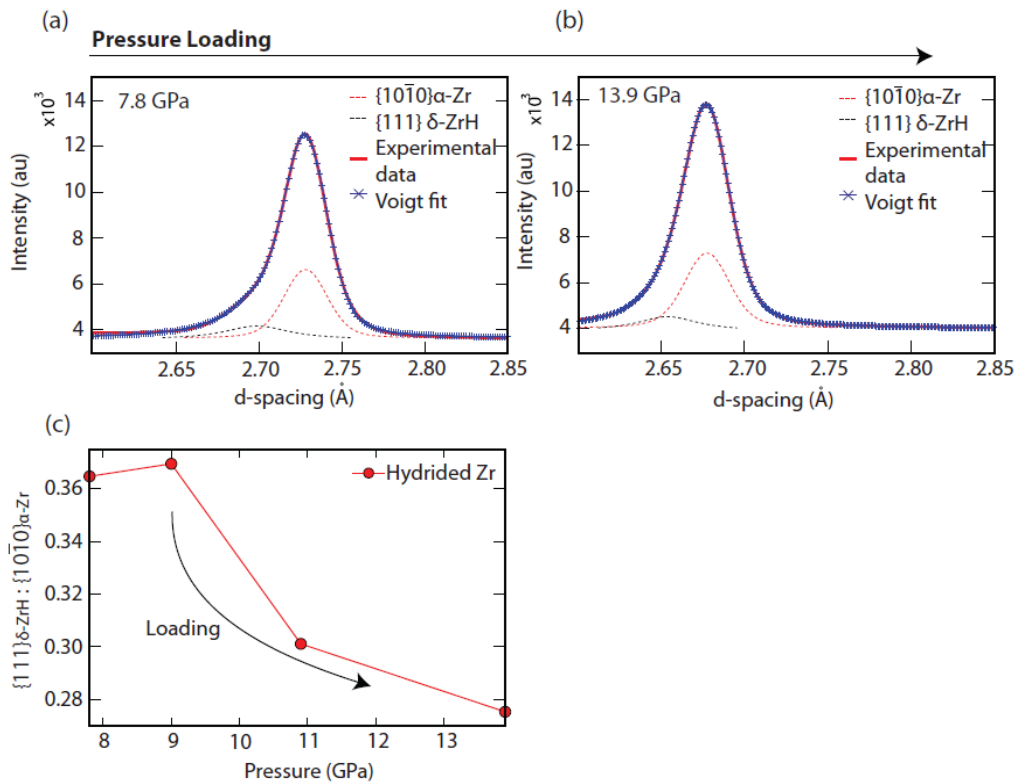

Figure 4. (colour online). (a-b) Variation in hydride phase $\{111\}_{\delta\text{-ZrH}}$ peak intensity relative to the neighbouring $\{10\bar{1}0\}_{\alpha\text{-Zr}}$ peak on loading in a Zircaloy-4 sample with δ phase already present, compressed in hydrogen PTM. (c) Change in $\{111\}_{\delta\text{-ZrH}}/\{10\bar{1}0\}_{\alpha\text{-Zr}}$ peak intensity ratio upon loading.

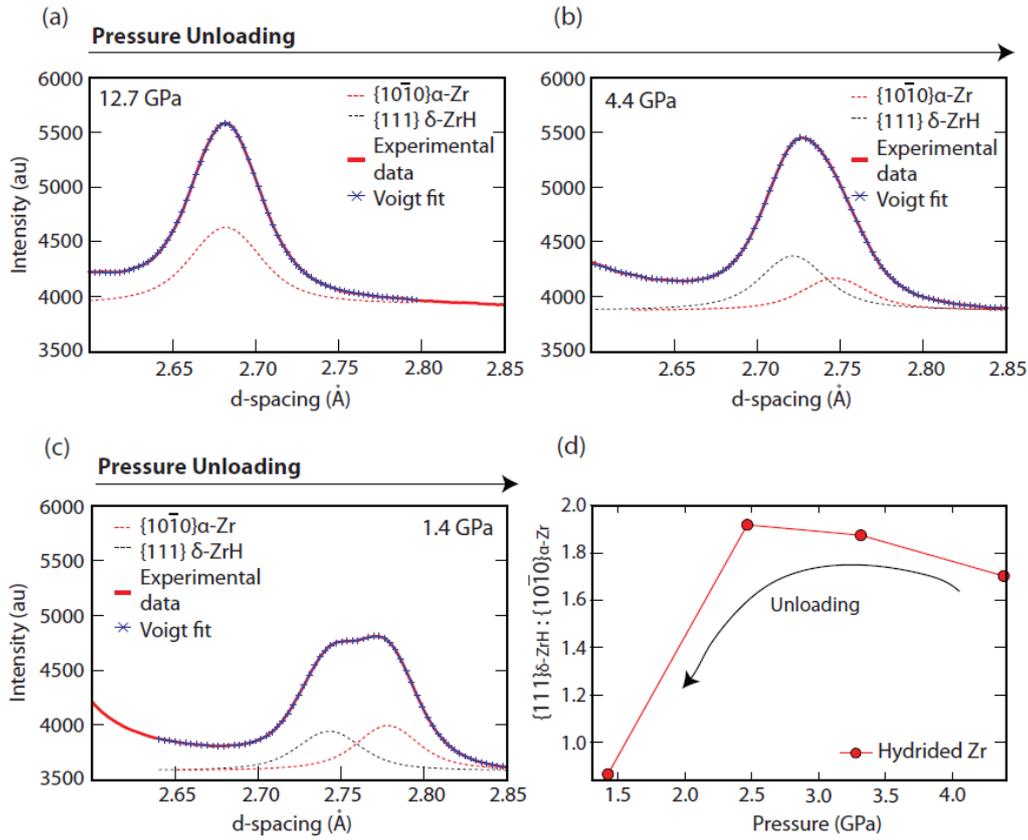

Figure 5. (colour online). (a-b) Variation in hydride phase $\{111\}_{\delta\text{-ZrH}}$ peak intensity relative to the neighbouring $\{10\bar{1}0\}_{\alpha\text{-Zr}}$ peak on loading in a Zircaloy-4 sample with δ phase already present, compressed in hydrogen PTM during the unloading phase (c). (d) Change in $\{111\}_{\delta\text{-ZrH}}/\{10\bar{1}0\}_{\alpha\text{-Zr}}$ peak intensity ratio upon unloading.

## 4. Discussion

The experimental observations are clear; hydrides in α-Zr dissolve under pressure and re-precipitate on the release of pressure, even in a hydrogen containing PTM. This occurs even when the pressure is raised high enough to precipitate the high pressure ω-Zr phase. Therefore, increasing pressure has the same effect on the Gibbs energy curves as increasing temperature, T, that is, it increases the solubility of hydrogen in α-Zr, see Figure 6. Thermodynamically, increasing T increases the entropy, S, contribution by –TS (where the configurational entropy is negative), and so it can be inferred that the entropy of δ-hydride increases more rapidly than that of α-Zr.

Given the lattice parameters of the three phases (see Table 3), the volume per zirconium atom can be calculated and indicates that ω-Zr has the highest density, followed by α-Zr, followed by the δ-hydride phase. This trend reproduces the observation that the application of hydrostatic compression favours (i) the ω phase, and (ii) α over δ; application of pressure results in dissolution of δ hydride. However, the results showed that ω only formed at 17-19 GPa, when in pure Zr it is known to form somewhere between 2-11.6 GPa [18-19], at 300 K, which implies that interstitial hydrogen retards the formation of ω. This is reasonable given that hydrogen expands the α-Zr lattice and is also a known β *bcc* phase (i.e. least dense phase) stabiliser, as shown by the Zr-H phase diagram [37]. In the case of our work hydrogen stabilises the α over the ω phase. This statement cannot be proved conclusively, however, as it does not take account the effect of the other alloying additions in Zircaloy-4, especially the oxygen, which is also known to stabilise α over ω as shown in titanium [38].

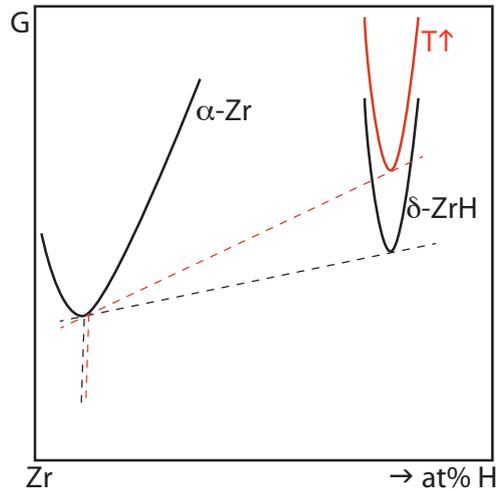

**Figure 6. Schematic depiction of the relative change in position of the α-Zr and δ-hydride Gibbs Energy curves as T increases (red), increasing the solubility of H in α-Zr and resulting in the dissolution of hydrides on heating. It is observed that increasing pressure has the same effect.**

Another mechanistic reason for the dissolution of hydrides on application of pressure is that the bulk modulus of the δ-phase is higher, 126-132 GPa [39-41], than that of the α-phase, 92 GPa [18]. This means that the hydride is less compressible than the matrix, meaning that the application of a given volumetric strain will require more energy in the hydride phase; this will also favour hydride dissolution, as observed.

As the thermodynamic system is alloyed and in a two-phase region, the transformation is continuous, with Figure 2(d) suggesting that 9 GPa is required to approximately halve the fraction of δ-hydride observed. Therefore, complete dissolution required around 18 GPa hydrostatic stress.

**Table 3. Cell parameters and volumes (per Zr atom). In the α phase, the interatomic spacing is taken as a; in δ as a/√2.**

| Phase | lattice parameters | | cell parameters | | V/nZr (Å³) | $(V_{phase}-V_\alpha)/V_\alpha$ (%) | Ref. |
|---|---|---|---|---|---|---|---|
| | a(Å) | c(Å) | $V_{phase}$: volume (Å³) | nZr (atoms per cell) | | | |
| α-Zr | 3.231 | 5.146 | 139.6 | 6 | 23.3 | 0.0 | 27 |
| ω-Zr | 5.039 | 3.316 | 102.4 | 6 | 17.1 | -26.6 | 18 |
| δ-ZrH | 4.78 | | 109.1 | 4 | 27.3 | 17.2 | 27 |

Recrystallised Zircaloy-4 has a yield stress, in uniaxial tension, less than 600 MPa, and therefore an upper bound for the tensile hydrostatic stress at a crack tip would be ~2.4 times the yield stress, ~1.4 GPa, as argued by McRae *et al.* [6] However, a (compressive) pressure of 1.4 GPa would only be sufficient to dissolve 8% of the maximum amount of hydride present.

MacEwan *et al.* [13] previously measured the molar volume of hydrogen in solution in α-Zr, concluding that the change in molar volume per dissolved H, $\overline{V_H}$, is around 1.67x10⁻⁶ m³ mol⁻¹, which is larger than for δ-phase, 1.4x10⁻⁶ m³ mol⁻¹. This results in the observation by dilatometry that hydride precipitation results in shrinkage of the material, which is consistent with the results above.

Relating this back to the mechanistic understanding of DHC appears that the consequence of this work is that at a crack tip loaded in tension, substantial hydride precipitation cannot occur, maybe only on the order of 0.5 mol%, on the assumption that hydrostatic compression and tension are analogous and reversible in sense.

**Conclusions**

The effect of pressure on the solubility of hydrogen in Zircaloy-4 at room temperature has been measured using synchrotron X-ray diffraction on small flake samples in a diamond anvil cell at pressures up to 20.9 GPa.

- An as-received sample placed in a hydrogen containing PTM resulted in the dissolution of δ hydride on the application of hydrostatic pressure, and the re-precipitation upon unload.
- α-Zr matrix, at its 300 K hydrogen solubility limit, of ~40wppm, in an inert PTM, did not precipitate hydride on the application of pressure, even at pressures sufficient to induce the formation of the high pressure ω-Zr phase of 17-19 GPa.
- A pre-hydrided sample containing hydride precipitates, in a hydrogen PTM, showed the dissolution of the hydride peak on the application of pressure. This was followed by ω-Zr transformation, and the re-precipitation of the hydride on pressure unload.

The results are consistent with the volume per Zr atom of the α, δ and ω phases, with the bulk moduli of α and δ, and with previous measurements of the hydrogen site molar volumes in the α and δ phases.

These results are then interpreted in terms of their implication for our understanding of the driving forces for hydride precipitation at crack tips, which are in a region of hydrostatic tensile stress on the order of 1.4 GPa in magnitude (but may be higher for, for example in cold worked or irradiation hardened material).

**Acknowledgements**
This work was carried out with the support of the Diamond Light Source (proposals EE11823, EE11658), EPSRC (grants EP/H004882/1, EP/K034332/1, EP/I003088/1) and EDF Energy (fellowship for MRW). We also gratefully acknowledge the provision of material by Rolls-Royce plc and useful discussions with Ted Darby. Dr John Plummer assisted with the synchrotron experimentation. We would like to acknowledge access to the high pressure hydrogen gas loading facility at the University of Edinburgh provided by Prof. E. Gregoryanz and the assistance of workshop staff at the University of Hull (N. Parkin) in the construction of the diamond anvil cells utilized in this work.  DS would like to acknowledge the provision of doctoral scholarships at the University of Hull and the University of Salford.

**Data Availability**
The raw and processed synchrotron data underlying these findings cannot be publicly shared at this time due to export control restrictions. Sharing of the data may be possible upon request to the corresponding author.

**References**
[1] R. Dutton, M. P. Puls, in: Effect of Hydrogen on Behavior of Materials, TMS-AIME, New York, 1976, pp. 512–525.
[2] R. Dutton, K. Nuttall, M. P. Puls, L. A. Simpson, *Metall. Trans.* **8A**, 1553–1562 (1977)
[3] Y. S. Kim, S. J. Kim, K. S. Im, *J. Nucl. Mater.* **335**, 387-396 (2004)
[4] Y. S. Kim., *Met. Mater. Inter.* **11**, 29-38. (2005)
[5] Y. S. Kim, S. B. Ahn, Y. M. Cheong, *J. Alloys Compd.* **429**, 221-226 (2007)
[6] G. A. McRae, C. E. Coleman, B. W. Leitch, *J. Nucl. Mater.* **396**, 130-143 (2010)
[7] M. P. Puls, *J. Nucl. Mater.* **393**, 350-367 (2009)

[8] M. P. Puls, *J. Nucl. Mater.* **399**, 248-258 (2010)
[9] Y. S. Kim, *J. Nucl. Mater.* **396** 144-148 (2010)
[10] Y. S. Kim, *J. Nucl. Mater.* **399**, 240-247 (2010)
[11] Y. S. Kim, *J. Nucl. Mater.* **399,** 259-265 (2010)
[12] C. E. Coleman, J. F. R. Ambler, *Scripta Mater.* **17**, 77-82 (1983)
[13] S. R. MacEwen, C. E. Coleman, C. E. Ells, *Acta Metall.* **33**, 753-757 (1985)
[14] M. S. Blackmur, M. Preuss, J. D. Robson, O. Zanellato, R. J. Cernik, F. Ribeiro, J. Andrieux, *J. Nucl. Mater.* **474**, 45-61 (2016)
[15] K. B. Colas, A. T. Motta, M. R. Daymond, J. D. Almer, *J. Nucl. Mater.* **440**, 586-595 (2013)
[16] H. E. Weekes, N. G. Jones, T. C. Lindley, D. Dye, *J. Nucl. Mater.* **478**, 32-41 (2016)
[17] H. E. Weekes, V. A. Vorontsov, I. P. Dolbnya, J. D. Plummer, F. Giuliani, T. B. Britton, D. Dye, *Acta. Mater.* **92**, 81-96 (2015)
[18] Y. Zhao, J. Zhang, C. Pantea, J. Qian, L. L. Daemen, P. A. Rigg, R. S. Hixson, G. T. Gray III, Y. Yang, L. Wang, Y. Wang, T. Uchida, *Phys. Rev. B* **71**, 184119 (2005)
[19] A. Dewaele, R. André, F. Occelli, O. Mathon, S. Pascarelli, T. Irifune, P. Loubeyre, *High Press. Res.* **36**:3, 237-249 (2016)
[20] A. McMinn, E. C. Darby, and J. S. Schofield. The Terminal Solid Solubility of Hydrogen in Zirconium Alloys. Zirconium in the Nuclear Industry: Twelfth International Symposium, ASTM STP 1354, pages 173–195, 2000.
[21] H. E. Weekes, Synchrotron X-ray diffraction investigations into the micromechanics of hydrides in Zircaloy-4, Ph.D. thesis, Imperial College London (2016)
[22] A. D. Chijioke, W. J. Nellis, A. Soldatov, I. F. Silvera, *J. Appl. Phys.* **98**, 114905 (2005)
[23] R. T. Howie, P. Dalladay-Simpson, E. Gregoryanz, *Nat. Mater.* **14**, 495 (2015)
[24] P. Dalladay-Simpson, R. T. Howie, E. Gregoryanz, *Nature* **529**, 63 (2016)
[25] S. Klotz, J-C Chervin, P. Munsch, G. Le Marchand, *J. Phys. D: Appl. Phys.* **42**, 075413 (2009)
[26] K. Trachenko V. V. Brazhkin, and D. Bolmatov, *Phys. Rev. E.* **89,** 032126 (2014)
[27] *Phase transformations of elements under high pressure*, E. Yu. Tonkov and E. G. Ponyatovsky, CRC Press (2005).
[28] R.J. Hemley, H. K. Mao, L. W. Finger, A. P. Jephcoat, R. M. Hazen, C. S. Zha . *Phys. Rev. B.* **42**, 6458.
[29] https://webbook.nist.gov/chemistry/fluid/
[30] O. Degtyareva, J. E. Proctor, C. L. Guillaume, E. Gregoryanz and M.Hanfland, Solid State Comms. **149**, 1583 (2009).
[31] B. Li, Y. Ding, D. Y. Kim, R. Ahuja, G. Zou and H.-K. Mao, PNAS **108**, 18618 (2011).
[32] T. Scheler, O. Degtyareva, M. Marques, C. L. Guillaume, J. E. Proctor, S. Evans and E. Gregoryanz, Phys. Rev. B **83**, 214106 (2011).
[33] T. Scheler, M. Marqués, Z. Konôpková, C. L. Guillaume, R. T. Howie and E. Gregoryanz, Phys. Rev. Lett. **111**, 215503 (2013).
[34] A.P. Hammersley, S.O. Svennson, M. Hanfland, A.N. Fitch and D. Hausermann, High Pressure Research **14**, 235 (1996).
[35] A. Steuwer, J. R. Santisteban, M. Preuss, M. J. Peel, T. Buslaps, M. Harada, *Acta Mater.* **57**, 145-152 (2009)
[36] J.M. Silcock, Acta Metall. **6**, 481-493 (1958).
[37] E. Zuzek, J. P. Abriata, A. San-Martin, F. D. Manchester, *Bull. Alloy Phase Diagr.***11**:4, 385-395 (1990)
[38] E. Cerreta, G.T. Gray III, A. C.Lawson, T. A. Mason, C. E.Morris. *J. Appl. Phys.* **100**, 013530 (2006)
[39] S. Yamanaka, K. Yamada, K. Kurosaki, M. Uno, K. Takeda, H. Anada, T. Matsuda, S. Kobayashi, *J. Alloys Compd.* **330-332**, 99-104 (2002)

[40] J. Blomqvist, J. Olofsson, A-M. Alvarez, C. Bjerken, *15th International Conference on Environmental Degradation of Materials in Nuclear Power Systems – Water Reactors* (2011)

[41] M. Christensen, W. Wolf, C. Freeman, E. Wimmer, R. B. Adamson, L. Hallstadius, P. E. Cantonwine, E. V. Mader, *J. Phys. Condens. Matter* **27**, 025402 (2015)